# Pathogen Infection Recovery Probability (PIRP) Versus Proinflammatory Anti-Pathogen Species (PIAPS) Levels: Modelling and Therapeutic Strategies

Sam-Shajing Sun

Department of Chemistry and Department of Physics
Norfolk State University
700 Park Avenue, Norfolk, Virginia, 23504, USA
(Correspondence:  ssun@nsu.edu)

**Abstract**

Current CoVID-19 pandemic is spreading rapidly worldwide, and it may become one of the largest pandemic events in modern history if out of control. It appears most of the SARS-CoV-2 virus infection resulted deaths are mainly due to dysfunctions or failures of the lung or multiple organs that could be attributed to host's immunodysfunctions particularly hyperinflammatory type disorders. In this brief review and study, a math model is proposed to correlate the Pathogen Infection Recovery Probability (PIRP) versus Proinflammatory Anti-Pathogen Species (PIAPS) levels within a host unit, where a maximum PIRP is exhibited when the PIAPS levels are equal to or around PIAPS equilibrium levels at the pathogen elimination or clearance onset. Based on this model, rational or effective therapeutic strategies at right stages or timing, with right type of agents (immuno-stimulators or immuno-suppressors), and right dosages, may be designed and implemented that are expected to effectively achieve maximum PIRP or reduce the mortality.

Current COVID-19 pandemic due to SARS-CoV-2 viruses have already spread around the globe and have resulted in over thirty thousand human deaths with twenty times more confirmed infections [1-2]. In addition to loss of human life, social and economic losses or effects could be significant. A number of earlier global pandemics occurred in human history can be attributed to pathogen infections [3]. Though there are differences among different pathogen induced infections, there were certain similarities among all pathogen infections. The pathogens here mainly refer to biological microorganisms such as viruses (including the new SARS-CoV-2 virus) and bacteria that can self-replicate in a biological host and can trigger or initiate a host immune system responses resulting in the production (clonal expansion) of anti-pathogen species (APS), including a series of proinflammatory anti-pathogen species (PIAPS). PIAPS here mainly refer to "double-edged sword" species such as certain white blood cells (WBCs) or their generated/related species, such as oxidative radical species and antibodies [4-6], cytokines [7-12, 18-20], *etc*. "Double-edged sword" refers to certain PIAPS that not only attack the pathogens but also attack host normal cells and tissues [4-12, 18-20].

Pathogen infection modeling could be very useful for understanding the infection mechanisms and processes, and for preventive or therapeutic strategies. However, most of the existing modeling works are mainly focusing on multiple host infection and transmittance statistics over time domain [13-17], very few modeling work provide insights on pathogen infection recovery probability (PIRP) over anti-pathogens species (APS), particularly over proinflammatory anti-pathogen species (PIAPS) that is the focus of this study.

A pathogen infection in a host may result in pathogen un-controlled growth if the host immune system is too weak, deficient, or dysregulated (including immunoparalysis and a serious immune deficiency syndromes) that could result in sepsis or septic shocks [20]. In a host with normal immune response system, as illustrated in Figure 1, the pathogen infection at time $t_0$ (end of incubation period) typically trigger a normal and efficient growth (clonal expansion) of immune system generated anti-pathogen species (APS at an initial level $x_0$) and ideally shall result in pathogen being eliminated/cleared at $t_e$ [13]. Once the pathogen is eliminated by the APS at $t_e$, the APS (including PIAPS) growth are expected to cease and remain at their equilibrium levels $x_e$. Certain APS (such as certain pathogen specific antibodies) are expected to remain at their equilibrium levels for certain period of time so the same pathogen infection can be prevented or inhibited (principle of vaccination), though antibody equilibrium level slow decay in long period of time are normal or expected [13-17].

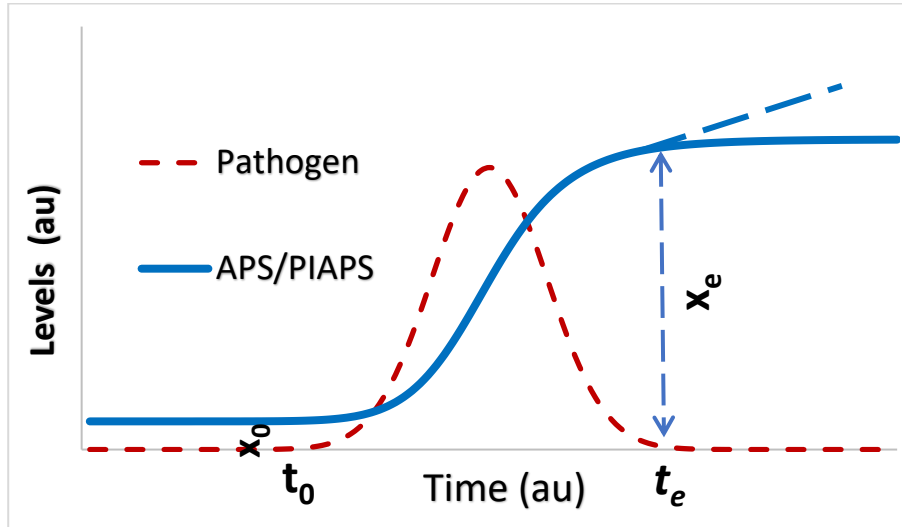

**Figure 1.** Schematic levels of pathogen (short dashed red curve) and host immune system generated anti-pathogen species (APS), including proinflammatory anti-pathogen species (PIAPS), for normal (solid blue curve) and abnormal (long dashed blue line, reflecting hyperinflammatory disorder) immune response reactions over time.

However, in certain immunodysfunction disorders, particularly certain hyperinflammatory disorders, such as cytokine release syndromes (CRS) or cytokine storm (CS) [4-12], macrophage activation syndromes (MAS) or macrophage-cytokine self-amplifying loop (MCSAL) [11], WBS proliferative disorders [4], certain PIAPS can grow out of control or not being efficiently dampened by the host anti-inflammatory species (*e.g.*, IL-10) even after $t_e$ where the pathogen may have been eliminated. It has been known that a number of PIAPS attack or damage normal or healthy cells resulting in tissue death (gangrene) and multiple organ dysfunctions or failures [2, 4-12, 18-20]. For this reason and for potential and practical therapeutic strategies, a Gaussian bell shaped normal distribution function $Y$ is proposed here to model the Pathogen Infection Recovery Probability (PIRP, or the survivability, counter to the mortality) versus the PIAPS levels $x$ (shown in Figure 2) and is exhibited with equation (1):

$$Y = \beta \, exp[-(x-x_e)^2/\alpha] \qquad (1)$$

where $\alpha$ parameter is proportional to the PIRP distribution peak full width at half maximum (FWHM) that affects the PIAPS level range width around PIRP maximum**.** During this range, PIAPS levels can significantly elevate PIRP as compared to other PIAPS range where PIRP remains relatively low. $\beta$ parameter represents a coupling factor of PIRP versus PIAPS levels, reflecting how significant or effective PIAPS level affects PIRP.

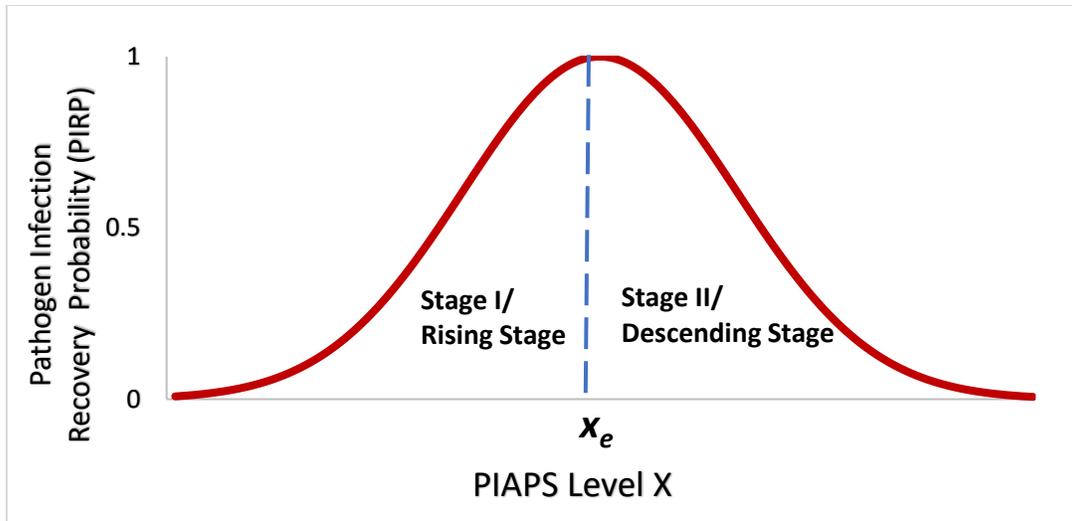

**Figure 2.** Scheme of Pathogen Infection Recovery Probability (PIRP) versus certain Proinflammatory Anti-Pathogen Species (PIAPS) levels based on equation 1.

Based on this model, the PIRP-PIAPS curve are divided into two stages: 1) Stage I or the PIRP rising stage corresponds to pathogen/APS evolution time period between $t_0$ to $t_e$ as shown in Figure 1: The PIRP of the pathogen infected host starts to rise as the host normal immune response generated APS (including PIAPS) are growing efficiently from initial levels of $x_0$ ($x_0$ can be zero for pathogen specific APS) and eventually approaching and maintaining at their equilibrium levels $x_e$ (blue solid line) where the pathogens are being eliminated or cleared. 2) Stage II or the PIRP descending stage: The PIAPS level further grow beyond their equilibrium levels $x_e$ as represented by the long dashed blue line (representing immunodysregulation such as hyperinflammatory disorders) [2, 4-12, 18-20], the PIRP descends presumably due to excessive PIAPS start to damage the normal or healthy tissues or organs. Eventually the PIRP could descend to a very low level due to heavy damages of tissues (particularly lung tissues) that could result in multiple organ failures [2, 7-12, 18-20].

Based on this model, the general therapeutic strategies for minimizing mortality is to achieve and/or to sustain maximum PIRP via a two-stage protocol as following: 1) In the stage I or the PIRP rising stage between $t_0$ and $t_e$, if the host has a normal immune response to the pathogen infection, the host's APS/PIAPS should grow efficiently toward their equilibrium (or saturation) levels $x_e$ where the pathogens are being eliminated or cleared. In this situation and stage, viral targeted therapies appear unnecessary except supportive therapies are needed for the following situations: a) If the host exhibits breath difficulty (dyspnea) or low blood oxygen level due to the liquids/mucous in the lungs (lung infections), then mechanical respiration ventilators and/or oxygen therapy may be utilized to prevent potential oxygen deficiency syndromes and related complications (hypoxemia and hypoxia); b) If the pathogen growth is out of control (such as in the cases of the host's certain immune deficiency syndromes), than either pathogen inhibitors/suppressors (if available) or APS boosters/enhancers (immuno-stimulators, including certain WBC therapies, antibody/immunoglobin therapies, interferon therapies, as well as therapies utilizing plasma and antibodies obtained from the convalescent patients) may be administered to minimize potential viral damage resulted complications, but the immuno-stimulations must be administered at the right time (in stage I before $t_e$), right type (APS/PIAPS boosters/enhancers instead of inhibitors/suppressors), and at the right dosages (*i.e.*, APS/PIAPS levels should be carefully monitored and controlled to be equal or close to their equilibrium or saturation levels $x_e$). 2) In the stage II or the PIRP descending stage after $t_e$, when the PIAPS levels are excessive or their growth are out of control (dysregulated), the most critical or essential therapeutic task in the post $t_e$ period or stage II shall be to promptly terminate or suppress the further growth of the PIAPS levels (immuno-suppressing, a variety of anti-inflammatory methods may be tested) at or nearby their equilibrium levels $x_e$, while pathogen inhibitors/suppressors may not be necessary at this time if the pathogens are eliminated. In case where the coupling of the host generated APS to the pathogen is very poor, *i.e.,* hyper-inflammation or cytokine storm has occurred and the pathogen level is still high, pathogen suppressors/inhibitors (if available), anti-inflammatory or non-inflammatory APS, as well as PIAPS suppressors may all be administered at this situation and in this stage but with carefully controlled dosages. Certain host immune system generated anti-inflammatory species (AIS, such as IL-10) may grow in order to counter the inflammation, but such anti-inflammatory response could be too slow and may eventually reduce the host PIAPS levels well below the equilibrium levels and may result the host to immunoparalysis [20]. A number of therapeutic PIAPS control (immunomodulation) efforts have been reported in recent years [4-12], however, the timing, type, and dosages of PIAPS suppressors/antagonists must be carefully monitored and controlled and this appears has not yet done, as PIAPS over-suppression or at wrong stage could result in delayed or incomplete pathogen elimination as well as vulnerability of host re-infection or secondary infections and related complications [20]. Finally, since the host's mental/psychological status or modes (fear including claustrophobia, anxiety, distress, depression, *etc*.) could trigger host's catecholamine/adrenaline production which in turn could boost APS/PIAPS levels, macrophage-cytokine self-amplifying loop MCSAL [11], and inflammations [21], and may result in mode-inflammation self-amplifying loop (MISAL), psychological counselling to the host thus also appear very important to improve host's PIRP. Precise, fast, convenient, and reliable protocols of measuring and monitoring pathogen and key "Double-edged sword" PIAPS levels are essential not only to validate this model, but to eventually utilize this model and its generated protocols for safe and effective therapeutic treatments of the infected hosts. Both pathogen and key PIAPS should be targeted as critical biomarkers ASAP.

As an example, in the case of COVID-19, while there appears lack of evidences of organ damages due to virus [22], excess levels or presences of macrophages, neutrophils, and inflammatory cytokines (such as Interleukin-6) were observed in multiple damaged organs in the autopsies and biopsies of the SARS-CoV-2 virus infected hosts [18]. Though APS/PIAPS boosters (such as interferon INF-alpha, gamma immunoglobulin, convalescent plasma collected from recovered patients) were recommended for COVID-19 infection treatments [18], based on this model, such treatments may best be used only for those hosts with deficient or very weak immune responses and should be administered in stage I. PIAPS suppression via a series of inflammation antagonists, or cytokine elimination via blood purification [18] appear useful for controlling CRS but they should be done after $t_e$ in the stage II, the PIAPS level control are extremely critical. Most importantly, the levels of SARS-coV-2 and key PIAPS levels (particularly IL-6, macrophages, neutrophils) at appropriate time intervals need to be measured and monitored precisely and closely in order to monitor and determine the virus growth, virus elimination onset time $t_e$ and the corresponding PIAPS equilibrium levels $x_e$. For COVID-19 infection, it appears many host's antibody lgG equilibrium level $x_e$ is about four times of its initial level $x_0$ [18]. An approach on controlling dysregulated interferon INF-I production in COVID-19 infection [19] appears potentially useful for validating or utilizing this model, again the interferon INF-I level control should be done after $t_e$ and the level should not be over suppressed well below $x_e$. Another example where this two-stage model might be applicable is the application of certain anti-oxidants (assuming Vitamins-C/E have such functions), where the anti-oxidant or radical scavengers appear needed only during stage II, this is because pathogen supressing oxidative radicals are actually needed in stage I. Finally, multiple host units may be utilized to obtain average values of all six parameters of this model ($t_0, x_0, t_e, x_e, \alpha, \beta$) for a particular host group, and the average values may be useful for therapeutic treatments of an individual host that is same or similar to the members of the group.

In summary, a bell shaped normal distribution function is proposed to model the Pathogen Infection Recovery Probability (PIRP) versus Proinflammatory Anti-pathogen Species (PIAPS) levels in a microorganism based pathogen infected host. Based on this model, therapeutic strategies should be based on two stages: In the first stage, treatments may not be necessary for most hosts with normal immune responses as PIRPs are expected to grow and remain at the maximum due to APS/PIAPS growing to and remaining at the equilibrium levels $x_e$ for certain periods, except supportive treatments are needed for oxygen deficiency syndromes. Hosts with weak or deficient anti-pathogen immune responses may need either pathogen suppressors or immuno-stimulators, however, timing, type, and dosages of both pathogen suppressors and immuno-stimulators are critical. In the second or the PIRP descending stage II due to PIAPS excessive or abnormal growth or levels, it is essential to control the PIAPS around their equilibrium levels $x_e$ via immuno-suppressors or inflammation antagonists. If pathogen levels are still high in stage II, then anti-inflammatory or non-inflammatory immuno-stimulators are desired. Again, timing, types, and dosages of therapeutic treatments are extremely critical depending on the PIRP stages and on pathogen/PIAPS levels. Precise and timely monitoring and controls of both pathogen and PIAPS levels are essential in order to fully characterize and utilize this model. Increased survivability or reduced mortality could be potential outcome if this or related models are fully developed, well characterized, and implemented after carefully designed and controlled clinical trials. For instance, for current COVID-19 infections, immunomodulation via timely and precise monitoring and level controls of key biomarkers (including the virus, IL-6, macrophages and/or neutrophils, oxidative radical species, IL-10, *etc*) appear essential to reduce the mortality.

**Acknowledgement**

The author wishes to thank his child's allergy specialist Dr. Kelly Maples for helpful discussions on cytokines/chemokines related hyper-inflammations in pathogen infections and food allergies, and to thank his colleague biochemistry Professor/Dr. Joseph Hall for helpful discussions on pathogen/anti-pathogens. The author also wishes to acknowledge his brother Mr. Honggang Sun on insightful discussions about "Happy Medium" doctrine of the Confucius philosophy that was established and taught in China for thousands of years. The author particularly wishes to acknowledge and thank his friend Dr. Donald Soles (MD, family medical practitioner) who has been overwhelmed in recent weeks trying to take care of many patients (including COVID-19 patients) and also graciously spent time to review and discuss with author on this article. The author wishes to express his deep appreciation and admiration to all medical staff (medical doctors, nurses, *etc*) worldwide who are risking their own lives trying to save lives of others in the current global humanity's battle against CoVID-19!

**Conflict of Interest**

The authors declare no any conflict of interest for publishing this article.

**References**

[1] COVID-19 Situation Report-55, World Health Organization (WHO), March 15, 2020. (https://www.who.int/emergencies/diseases/novel-coronavirus-2019)

[2] Chen Y, Liu Q, Guo D. "Emerging coronaviruses: Genome structure, replication, and pathogenesis". *J Med Virol.*, **92:**418–423 (2020) (https://doi.org/10.1002/jmv.25681)

[3] https://en.wikipedia.org/wiki/Pandemic

[4] Leslie, M., *"The body's dangerous defenders", Science* **367** (6482), 1067-1069 (2020) (DOI: 10.1126/science.367.6482.1067).

[5] Vandenhaute, J., Wouters, C., Matthys, P., "Natural Killer Cells in Systemic Autoinflammatory Diseases: A Focus on Systemic Juvenile Idiopathic Arthritis and Macrophage Activation Syndrome", *Front. Immunol.,* 15 January 2020. (https://doi.org/10.3389/fimmu.2019.03089)

[6] Wesemann D., Nagler, C., "Origins of Peanut Allergy-Causing Antibodies", *Science* **367** (6482), 1072-1073 (2020) (DOI: 10.1126/science.aba8974).

[7] Hay, K., *et al.*, "Kinetics and biomarkers of severe cytokine release syndrome after CD19 chimeric antigen receptor–modified T-cell therapy", *BLOOD*, **130**, 21 (2017) (DOI 10.1182/blood-2017-06, 793141)

[8] Liu, Q., Zhou, Y., Yang, Z., "The cytokine storm of severe influenza and development of immunomodulatory therapy", *Cellular & Molecular Immunology*, **13**, 3–10 (2016) (doi:10.1038/cmi.2015.74)

[9] Tisoncik JR, Korth MJ, Simmons CP, Farrar J, Martin TR, Katze MG. "Into the eye of the cytokine storm", *Microbiol Mol Biol Rev*. **76(1):**16–32 (2012) (doi:10.1128/MMBR.05015-11)

[10] Luo C, Liu J, Qi W, *et al*. "Dynamic analysis of expression of chemokine and cytokine gene responses to H5N1 and H9N2 avian influenza viruses in DF-1 cells". *Microbiol Immunol.*, **62(5):**327–340 (2018) (doi:10.1111/1348-0421.12588).

[11] Staedtke, V., *et al.*, "Disruption of a self-amplifying catecholamine loop reduces cytokine release syndrome", *Nature*, **564**, 273 (2018) (https://doi.org/10.1038/s41586-018-0774-y).

[12] Teijaro, J., Walsh, K., Rice, S., Rosen, H., Oldstone, M., "Mapping the innate signaling cascade essential for cytokine storm during influenza virus infection". *Proceedings of the National Academy of Sciences*, **111 (10)** 3799-3804 (2014) (DOI: 10.1073/pnas.1400593111)

[13] Gulbudak, H and Browne, C., "Infection severity across scales in multi-strain immuno-epidemiological Dengue model structured by host antibody level", arXiv:1912.08305 [q-bio.PE], December 17, 2019.

[14] Barbarossa, M. and Rost, G., "Immuno-epidemiology of a population structured by immune status: a mathematical study of waning immunity and immune system boosting", arXiv:1411.3195 [math.DS], November 14, 2014.

[15] Uekermann, F., Sneppen, K., "Cross-immunization model for the extinction of old influenza strains". *Sci Rep.,* **6,** 25907 (2016). (https://doi.org/10.1038/srep25907).

[16] Siettos, C. & Russo, L., "Mathematical modeling of infectious disease dynamics", *Virulence,* **4:4**, 295-306 (2013) (DOI:10.4161/viru.24041)

[17] Zhu X, Kranse R, Bul M, Bangma CH, Schröder FH, Roobol MJ. "Overestimation of prostate cancer mortality and other-cause mortality by the Kaplan-Meier method". *Can J Urol*. **20(3**):6756–6760 (2013) (PMID:23783043).

[18] "Novel Coronavirus Pneumonia Diagnosis and Treatment Plan (Provisional 7th Edition), National Health Commission of China (NHCC), March 4, 2020. (https://www.chinalawtranslate.com/en/coronavirus-treatment-plan-7/).

[19]   Deng, X., Yu , X., Pei, J., "Regulation of interferon production as a potential strategy for COVID-19 treatment", arXiv:2003.00751 [q-bio.MN], March 2, 2020.

[20]   Monneret, G., Venet, F., Pachot, A., Lepape, A., "Monitoring Immune Dysfunctions in the Septic Patient: A New Skin for the Old Ceremony". *Molecular medicine* (Cambridge, Mass). **14 (1-2**), 64-78 (2008). (DOI: 10.2119/2007-00102.Monneret).

[21]   Berk, M., Williams, L.J., Jacka, F.N. *et al.* "So depression is an inflammatory disease, but where does the inflammation come from?" *BMC Med* **11,** 200 (2013). (https://doi.org/10.1186/1741-7015-11-200).

[22]   Xu, Z. *et al.*, "Pathological findings of COVID-19 associated with acute respiratory distress syndrome", *The Lancet Respiratory Medicine* (online), February 18, 2020 (https://doi.org/10.1016/S2213-2600 (20)30076-X).